# Early Phishing

Koceilah Rekouche  krekouche@pushstart.info

The history of phishing traces back in important ways to the mid-1990s when hacking software facilitated the mass targeting of people in password stealing scams on America Online (AOL). The first of these software programs was mine, called AOHell, and it was where the word phishing was coined. The software provided an automated password and credit card-stealing mechanism starting in January 1995. Though the practice of tricking users in order to steal passwords or information possibly goes back to the earliest days of computer networking, AOHell's phishing system was the first automated tool made publicly available for this purpose.[1] The program influenced the creation of many other automated phishing systems that were made over a number of years. These tools were available to amateurs who used them to engage in a countless number of phishing attacks. By the later part of the decade, the activity moved from AOL to other networks and eventually grew to involve professional criminals on the internet. What began as a scheme by rebellious teenagers to steal passwords evolved into one of the top computer security threats affecting people, corporations, and governments.

## 1.    Prelude to Automation

In 1994, there was a small community of people on America Online who were self-identified computer hackers. I became part of this community, at age 16, and was known under the pseudonym "Da Chronic." Many of us were friends and we lived in various parts of the country. One constant requirement for all of us was maintaining our free and anonymous access to the service. At some point during the summer of that year a fellow hacker, Dave Lusby (Soul Crusher), explained to me the method he had developed to trick AOL members by posing as an employee of the service. In this way he would gain access to their accounts and credit cards. Similar to today's phishing schemes, it involved tricking someone into trusting you with their personal information. In this case, a person who had just logged on to cyberspace for the first time would be fooled into giving up their password or credit card information. Stealing accounts in this manner became my primary way of maintaining access to the service.

This early scheme was the basis for the automated tools that would be developed in the following year. One of us named it "fishing," the term that I later used in AOHell along with a change of spelling to "phishing."

---

[1] Though I do not know of any cases, it is plausible that some had previously developed tools for their own use in order to automate their scams, such as shell scripts that spam email (bait) to a large number of users at one time.

# 2  Koceilah Rekouche

## 1.1  How it Worked

AOL gave little or no warning to their customers about password and credit card scammers in 1994 and for most of 1995. The "New Member Lounge" chat rooms were our primary target areas since they were the rooms in which new subscribers were placed during their first few logins. Many people occupying these areas had only a few minutes of experience online, so they were some of the most vulnerable targets an attacker could want.[2] The method was as follows:

1. Obtain an anonymous AOL account by creating one using a fake bank account number or credit card, or use an account that was stolen in a previous attack.
2. Create a screen name on the account that appears official (e.g. BillingDept).[3]
3. Write the "bait" message which will explain to users the need for us to "verify" their passwords or billing information. For example: "*Hi, this is AOL customer service. Due to a problem with our records, we need you to reply to this message with your current password in order to avoid being disconnected.*"
4. Locate a New Member Lounge chat room and open its occupant list.
5. Send a private message containing the bait to each person in the room.

Steps 4 and 5 are repeated until the account is terminated by AOL security staff (typically less than five minutes).

## 1.2  Motivation & Weak Security

Though one could already obtain free access to AOL by creating accounts using fake credit card numbers, these accounts would only last a short time before AOL closed them when it attempted to verify the billing information. This made stolen accounts much more valuable as they were being paid for by legitimate subscribers.[4] Stolen credit card information was valuable to us for creating accounts on other online services that required verification of billing before the account was activated, such as The Sierra Network, CompuServe, GEnie, and Delphi.

---

[2] Since this is a technical description of early phishing, I am using the term "target" to refer to the people being victimized. The desire is not to depersonalize the nature of these attacks as the real lives of people were (and are) affected by phishing scams.

[3] Creating screen names that would appear to belong to AOL employees was hampered by AOL's efforts to block the creation of screen names containing certain key words. Some examples are names containing 'AOL', 'billing', 'TOS', 'account', etc. To circumvent this, we would use "visual deception": slightly misspelling the word while keeping it similar enough so that targets would not notice (e.g. spelling 'AOL' with a zero or 'billing' with two capital Is instead of lower case Ls).

**3  Koceilah Rekouche**

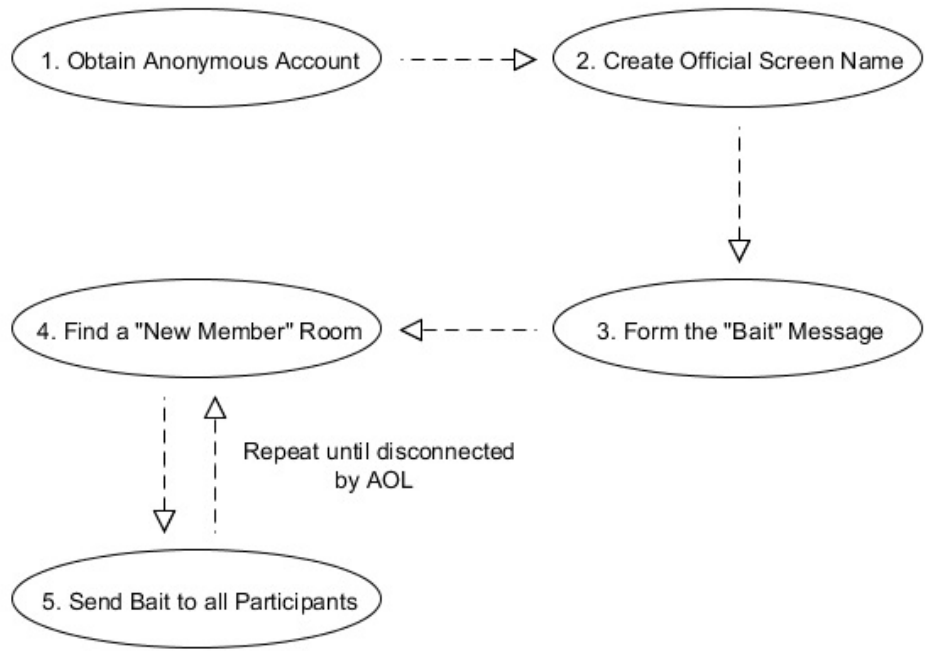

Figure 1: Password and Credit Card Scam Process

---

[4] These accounts were later referred to as "phishes." While a stolen account was more valuable than an account made using fake billing information, there were still problems with using them. AOL would allow only one screen name to be online at one time. If a user tried to login to his account while a screen name associated with it was already online, he would receive a message explaining the situation and informing him to call AOL's customer service. This is how many subscribers discovered that they had been duped by someone pretending to be an AOL employee, and this is how most phished accounts were "lost." The most valuable phishes were naturally ones belonging to subscribers who rarely signed in to their accounts. During this period AOL charged $3.50 per hour of online time, so many subscribers would discover what had happened after receiving a large bill.

AOL's lax verification system gave us free and anonymous access to their service from which we could target legitimate customers, a security weakness that would later become crucial in allowing massive numbers of people to engage in automated attacks. Much of that could have been prevented if AOL had implemented an automatic billing verification system similar to the ones found on the much smaller online services.[5]

Throughout 1994, only a small number of us engaged in the "fishing" scheme–Lusby and I, and perhaps a few others. This would change the following year when AOL hacking software started to become popular, and our small underground community grew enormously due to both the software and AOL's growing subscriber base. I started the development of this software in late 1994 with another programmer, and by early 1995 I had developed an automated phishing system.

## 2.     Automatic Phishing

Figure 2 shows the configuration screen of the attack system included with AOHell. Here the person chooses from built-in bait messages or creates his own. Messages are automatically sent to AOL subscribers asking for their passwords or credit card information. So that a novice could use the system, a Help button was prominently displayed which provided information on the concept and explained how to use the system to effectively steal passwords and credit cards.

The first recorded uses of the term phishing are found here in the program's two interface windows as well as in the documentation. (Both spellings are used; with an "f" and with "ph.")

---

[5] However, AOL was much more concerned with unhindered growth rates even if the growth was not completely real. AOHell would later provide a function that would create these pre-verified accounts automatically for the user. This resulted in the creation of tens of thousands of fake accounts, many of which were used for stealing real accounts. This could have been prevented by a simple billing verification system. Even at this early time such systems were common among online services and merchants. AOL announced its position on its weak security when Vice President Kathy Ryan stated that "[AOL understands] that our aggressive distribution of both software and certificates can result in 'throwaway' accounts. We have made the business decision that the benefits in this case outweigh the disadvantages." AOL was also reluctant to warn its customers about the threat of phishers during the first few months of AOHell-related phishing attacks. See David Cassel, "Hackers, Netscape, Death of AOL?," *AOL Watch Mailing List* (2 April 1999), (http://www.aolwatch.org/list/0101.html).

# 5 Koceilah Rekouche

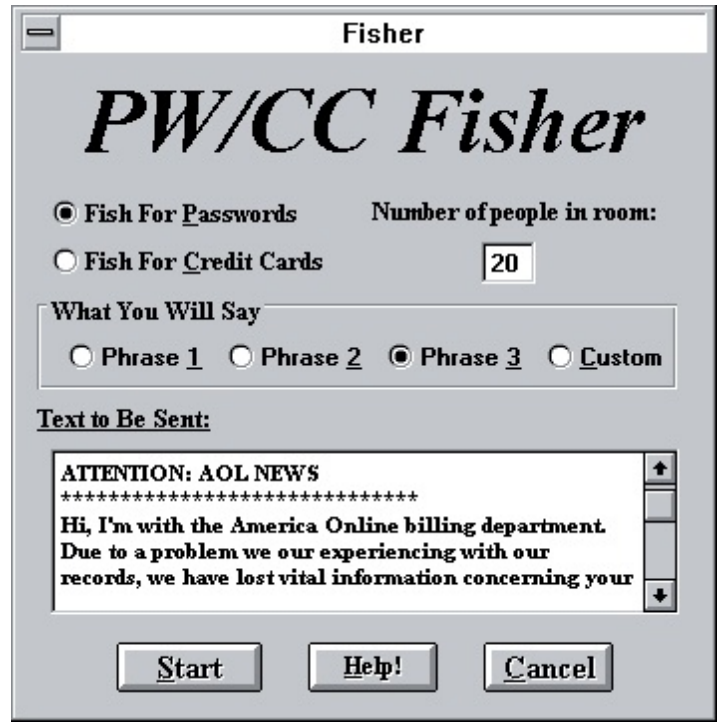

Figure 2: AOHell's Phishing Configuration Screen

Figure 3 is an accurate re-creation of a phishing attack in progress from the phisher's point of view. Messages are quickly sent to occupants in a chat room. The user waits for each person to be sent a message. He is then prompted to locate the next chat room to phish. The phisher targets room after room until the account is terminated by an AOL employee. He makes no actual communication with the targets himself. Incoming replies appear only briefly on the screen and are logged for later retrieval.

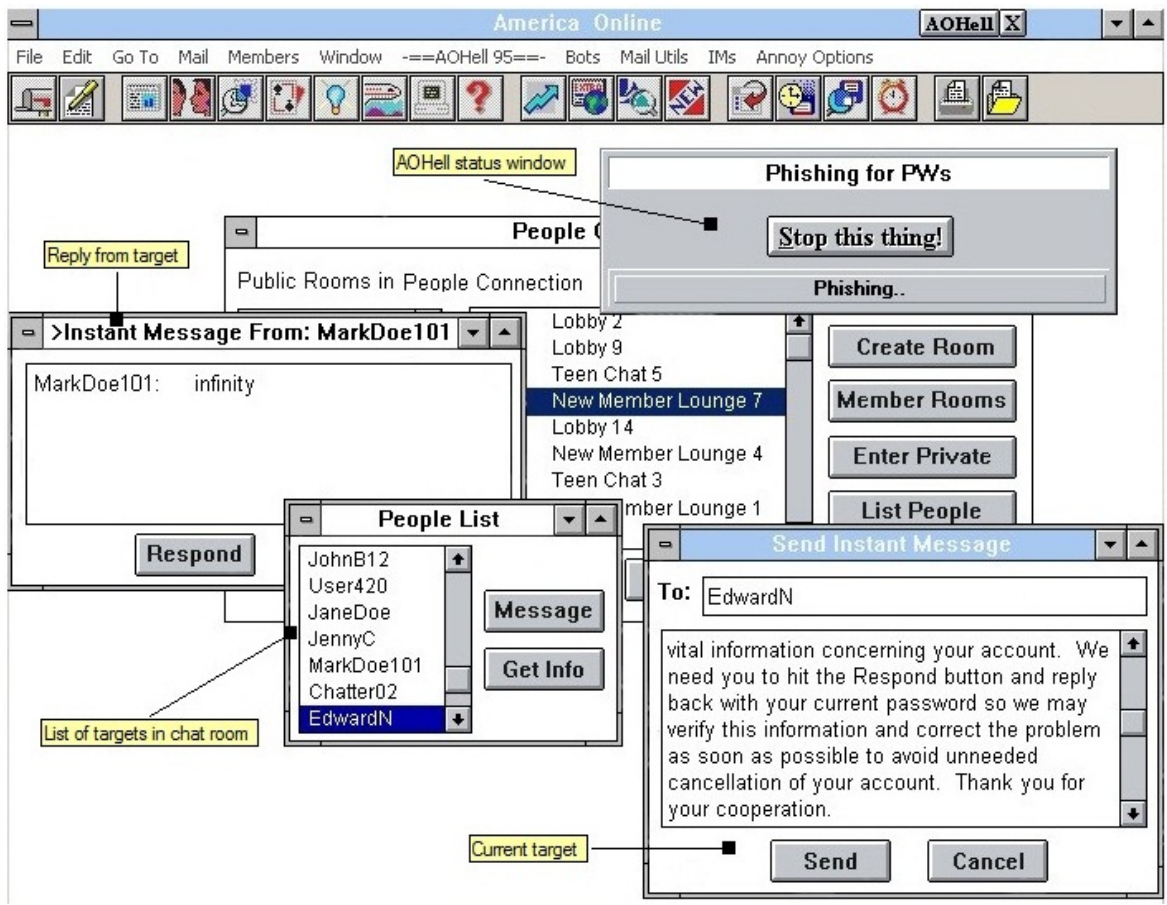

Figure 3: Automated Phisher in action on Windows 3.1 and AOL 1.1

## 2.1    Phishing Made Easy

AOHell performed many functions useful to hackers and software traders and gained widespread usage very quickly. The program included as one of its features an automated phishing mechanism starting in January 1995. One of my objectives in making the program was that it should be used by a large range of people, from experienced users to people who had very little computer knowledge. The design approach was to assume that the user knew nothing about a particular procedure.[6] The phishing configuration screen (Figure 2) offered the user detailed documentation through a Windows Help file that explained what phishing was, what the person would gain by doing it, and how to get the best results from the automated system, such as choosing appropriate screen names. An easy to use graphical interface was provided along with the detailed instructions.

The user-friendly aspects of the program were crucial to the process in which phishing became such a widespread phenomenon. Not only could hackers and others with experience use the system, but anyone who was willing to follow some instructions and press a few buttons could phish. A very large number of people, who would not have otherwise engaged in such activity, became efficient password and credit card thieves with little knowledge and work required. These individuals would be aptly referred to as "script kiddies" in today's vocabulary.

## 2.2    Operating the Software

Like the scheme from the year prior, the operation of the automated system was not complicated. The software included six customizable bait messages, three each for password and credit card-phishing. The user would decide what he wanted to phish for, open a chat room occupant list, and start the phisher. The program would then send the bait to the people in that room one by one. After the last person had been sent a message, the user would open another room's occupant list and start the system again. Incoming messages from targets were removed from the screen, after being logged, to prevent interference with the process.

Once the automated system was started, there was little for the user to do except to periodically redirect AOHell to new chat rooms and wait to be disconnected by AOL security. AOL was usually diligent in disconnecting the phisher quickly. This was partly because the software would inadvertently send bait messages to AOL employees sitting in the chat rooms. Both the automated and the earlier manual attacks would typically last less than five minutes before the phisher's account was terminated. Since the time available for an attack was so short, automation provided a huge increase in results. With the aid of the software, perhaps a dozen chat rooms with up to 23 people each (200-300) could be targeted before the phisher was disconnected from the service.

---

[6] A major goal in writing AOHell was to gain a user base not just within AOL's hacking community but, more importantly, to get users from outside this community and thus increase its size by recruiting and educating new people. This was extremely successful as the popularity of AOHell and similar programs were largely responsible for growing the warez, hacking, and programming communities to a point where they reached thousands of participants. For each new release, and periodically in between releases, I would spam a copy of the program, along with a layman's description of the things that it could do, to every person in the Teen Chat rooms. This was a very effective way of getting new people to use the program as email spamming had not yet come about. Phishing was one component of the software, but most AOL teenagers were drawn by the other advertised functions such as the ability to "punt" their friends offline or the ability to scroll ASCII art in the chat rooms.

## 2.3   Big Rewards with No Risk

Unlike today's phishing schemes which operate in an environment that has over fifteen years of phishing history, these early attacks had a very good success rate against their unsuspecting targets. The high success rate was also due to the unique operation of the New Member Lounge areas, as previously mentioned. On busy nights, and on occasions when AOL staff were not so diligent, many passwords or credit cards could be phished in a single session.

Because the phishing was done on fake or stolen accounts, there was no substantial risk to the attacker of being caught by AOL or the police.[7] AOL relied on outside networks such as SprintNet for their telephone-modem connectivity, and these networks did not provide AOL with the incoming telephone number of the person logging in, say, through Caller ID. AOL later created or purchased its own modem network called AOLNet, but apparently it too did not retrieve the incoming telephone number.

## 3.   Phishing Evolves

Over the course of 1995, AOHell's phishing system was used in thousands of attacks. They continued and multiplied when software with similar functionality became prevalent. I stopped development of AOHell that September, but by this time there were already dozens of similar programs, many of which contained innovations and improvements. Many hundreds of these programs were developed over the course of several years, and they helped to create and support large underground communities on AOL. Phishing continued to be a major activity on AOL throughout.

Most of those involved with phishing on AOL were teenagers who were not well organized but who took part in loose communities. Software trading, developing underground software, and phishing were done for fun and peer acceptance. During the later part of the 1990s, people within these communities brought phishing to other services on the open internet, such as instant messaging and video game networks. There is much more to say about the period of widespread phishing on America Online, the large number of people involved, the diversity of the underground communities, and the motivations and histories behind them.

---

[7] Many years later, some AOL phishers were caught and prosecuted such as Helen Carr.

## 9  Koceilah Rekouche

Since the transition period from AOL to the internet, phishing, while never morally correct, has largely become an activity of organized criminals done for monetary gain. Beginning roughly in the early 2000s, phishers have used increasingly sophisticated methods to steal money from people and financial institutions. While an early phishing attack on AOL targeted a small number of people in an attempt to steal passwords or credit card information, a modern phishing attack on the internet can involve millions of targets in an attempt to steal entire bank accounts, or worse. Estimates of the annual costs of phishing to people and businesses range from tens of millions to billions of dollars per year.

Phishing would still exist today had it not first appeared on a massive scale on America Online in 1995, though it would be called something else and would have different forms and character. A worthwhile project would be to examine the evolution of phishing by tracing in detail its departure from AOL to other online networks. This would include an examination of many of the hundreds of software tools available from 1995 to the early 2000s when phishing became widespread on the internet.